\UseRawInputEncoding % for arxiv 

\documentclass[aps,showpacs,onecolumn,notitlepage,floatfix,amsmath,amssymb,nofootinbib]{revtex4-1}

\usepackage{amssymb}
\usepackage{amsmath}
\usepackage{epsfig}
\usepackage{graphicx}
\usepackage{subfig}
\usepackage{color}
\usepackage{latexsym}
\usepackage{bm,latexsym}
\usepackage{mathrsfs}
\usepackage{float}
\usepackage{pbox}

\usepackage[usenames,dvipsnames]{xcolor}
\definecolor{bostonuniversityred}{rgb}{0.8, 0.0, 0.0}

\selectfont{}
\begin{document}
%%%%%%%%%%%%%%%%%%%%
%\onehalfspacing
%%%%%%%%%%%%%%%%%%
\title{{\bf Transforming singular black holes  
into regular black holes sourced by nonlinear electrodynamics.}}

\author{Pedro Ca\~nate$^{1,~\!\!2}$}
\email{pcannate@gmail.com}

\author{Santiago Esteban Perez Bergliaffa$^{1}$}
\email{sepbergliaffa@gmail.com}

\affiliation{$^{1}$Departamento de F\'isica  Te\'orica, Instituto de F\'isica, Universidade do Estado do Rio de Janeiro,\\ Rua S\~ao Francisco Xavier 524, Maracan\~a \\
CEP 20550-013, Rio de Janeiro, Brazil.\\
$^{\!^2}$Programa de F\'isica, Facultad de Ciencias Exactas y Naturales, Universidad Surcolombiana, Avenida Pastrana Borrero - Carrera 1, A.A. 385, Neiva, Huila, Colombia.}

\begin{abstract}
We present a procedure that transforms a singular, asymptotically flat, static and spherically symmetric black hole 
into a regular
black hole spacetime. 
The regular black hole is a solution of General Relativity theory coupled to nonlinear electrodynamics (NLED), even if the original metric is not  a vacuum, or electro-vacuum, solution of General Relativity. 
\end{abstract}


% 04.20.Jb Exact solutions
% 04.50.Kd Modified theories of gravity
% 04.50.?h Higher-dimensional gravity and other theories of gravity
% 04.40.Nr Einstein-Maxwell spacetimes, spacetimes with fluids, radiation or classical fields

\maketitle

\section{Introduction}
%%%%%%%%%%%%%%%%%%%%%%%%%%%%%%%% 

Black hole
spacetimes 
are solutions of a given gravitational theory 
that display many interesting features, and 
are realized in Nature, according to the observational evidence presented for instance in \cite{L&V,EHT}\footnote{See 
\cite{Wald,TWHs,BHs, Rubio2018} for black holes (BHs) in
general relativity (GR) and modified gravity (MG) theories}. 
A fundamental trait of BHs is the presence 
of at least one event horizon, which may or may not hide a singularity.
Among the several definitions of singular spacetimes, we can cite the following two:
(1) a spacetime is singular if is timelike or null geodesically incomplete, and (2) a spacetime is singular if any of the curvature invariants (either constructed from the Riemann tensor, such as $R \equiv R^{\alpha}{}_{\alpha}$, $R_{\alpha\beta}R^{\alpha\beta}$, $R_{\alpha\beta\mu\nu}R^{\alpha\beta\mu\nu}$, or 
formed by polynomial expressions in covariant derivatives of the Riemann tensor) diverge in some region of the spacetime. 
Specifically, in the case of the
exact
black hole spacetimes 
in the GR context
(namely, the Kerr-Newman geometry and its particular cases), these two notions of singularity coincide (for details, see section 5.1.5 of \cite{Senovilla}).

Since singular spacetimes possess several undesirable features (see for instance
\cite{Romero2013}), ways to avoid them both in classical and quantum contexts
have been frequently presented in the literature. 
In particular, classical
regular solutions
are worth studying
due to the absence of a
complete
microscopic theory for the gravitational field\footnote{We shall be concerned here with classical regular black hole solutions. For nonsingular cosmological models, see for instance 
\cite{Novello2008}.}. 
Regular, asymptotically flat and spherically symmetric black hole spacetimes have been found 
by adopting 
different
sources for Einstein's equations, such as
charged matter
\cite{Lemos2011},
minimally coupled scalar fields with arbitrary
potentials and negative kinetic energy 
\cite{Bronnikov2006}, 
and
nonlinear electrodynamics (NLED) (see
\cite{Bardeen1968, Ayon1998,
 Ayon1999,Dymnikova2003,
 Dymni2004, Balart2014}, among others)\footnote{For a recent review of the regularity conditions in the latter type of solutions, see
\cite{Alberto_Gustavo}.}$^,$\footnote{Combinations of such sources have also been
employed, see for instance
\cite{Bronnikov2022}.}. In the latter case, which we shall call GR-NLED, the known charged singularity-free black hole solutions are either purely electrical, with  electromagnetic invariant $\mathcal{F} \sim -{\bf E}^{2}$ (where ${\bf E}$ denotes the electric field vector), or purely magnetic with $\mathcal{F} \sim {\bf B}^{2}$ (where ${\bf B}$ denotes the magnetic field vector). 
In addition to being regular, such solutions satisfy the weak energy condition and  do not admit a Cauchy surface. Hence, they do not contradict the Penrose singularity theorem \cite{Penrose} 
which,
to show the (null) geodesically incompleteness 
of a spacetime, 
assumes 
the fulfillment of the weak energy condition (WEC), the existence of a noncompact Cauchy surface, and the existence of a closed trapped surface.

Most of the regular solutions in the framework of GR-NLED existing in the literature
(as well as most of the solutions with other sources mentioned above)
have been obtained starting from a regular BH metric, from which the source follows via the field  equations.
 In all cases, it was verified that the Lagrangian corresponding to the source 
is such that the minimal requirements for the source to be physically acceptable are met. Such a ``reverse-engineering´ process'' has been successfully used also in the construction of wormhole solutions
\cite{WEC}, and to obtain regular solutions in theories other than GR (see for instance 
\cite{Olmo} for Born-Infeld gravity, and \cite{Rodrigues}
for $f(R)$ gravity).

The main goal of the present work is to show that actually all regular, asymptotically flat and spherically symmetric black hole solutions of
GR-NLED can be obtained by a procedure which basically consists in cutting out 
from a singular, zero charge BH metric
a spherical region
that includes the singularity, 
in such a way that the radius of the 
excised region is related to the charge of the regular black hole thus obtained. 
Hence, the procedure  transforms a singular  uncharged
static, spherically symmetric and asymptotically flat (SSS-AF) BH spacetime into a a regular charged SSS-AF BH spacetime in the framework of GR-NLED.

Let us remark that the procedure that leads to the removal of the singularity  presented here can be used to reproduce known solutions of the
GR-NLED system, and also to generate new solutions. It will be shown that the new solutions satisfy a theorem regarding regular BHs
with magnetic sources in NLED
presented in \cite{Bronnikov2000}. In fact, their existence suggest that there is a theorem that is dual to the one presented in \cite{Bronnikov2000}, as discussed below. 

The paper is organised as follows. In Section
\ref{sssr} we show how a static, spherically symmetric and asymptotically flat (SSS-AF) spacetime
of the 
black hole type in the which the only physical singularity  is localized at the origin, can be modified to get a new geometry representing a regular black hole spacetime.
In section \ref{BBinGR-NLED} the existence of theses kind of  
spacetime geometries as exact solutions of GR-NLED is explored, and examples of purely magnetic, purely electric, and hybrid solutions are displayed. We close with our conclusions in Section \ref{concl}\footnote{In this paper we use units where $G = c =\hbar = 1$, metric signature $(-+++)$ is used throughout.}.

%%%%%%%%%%%%%%%%%%%%%%%%%%%%%%%%%%%%%%%%%%%%%%%%%%%%%%%%%%%%%%%%%%%%%%%%%%%%%%%%%%%%%%%%%%%%%%%%%%%%%%%%%
\section{Removal of the singularity of a static, spherically symmetric black hole}
\label{sssr}
Let us present next a simple way to modify a given SSS-AF black hole 
spacetime (which is not necessarily a vacuum, or electro-vacuum,
solution of Einstein's equations) to obtain a new geometry of the SSS-AF regular black hole type, which solves Einstein's equations with a given NLED theory as a source. 
The starting metric, given by
$\boldsymbol{ds^{2}} = g_{_{tt}}\boldsymbol{dt}^{2} + g_{_{rr}}\boldsymbol{dr}^{2} + r^{2}\boldsymbol{d\Omega^{2}}$ being $\boldsymbol{d\Omega^{2}}= \boldsymbol{d\theta^{2}} + \sin^{2}\theta \boldsymbol{d\varphi^{2}}$ the line element of a %two-dimensional 
unit 2-sphere,
has a physical singularity at $r=0$, and is such that $g_{_{tt}}g_{_{rr}} =-1$.
For the
singular SSS-AF BH metric we shall adopt the
following {\it{Ansatz}}:
\begin{equation}\label{BH_structure}
\boldsymbol{ds^{2}}
= - \left( 1 - \frac{2 \mathcal{M}(r)}{r} \right)\boldsymbol{dt^{2}} + \left( 1 - \frac{2 \mathcal{M}(r) }{r} \right)^{-1} \boldsymbol{dr^{2}} + r^{2}\boldsymbol{d\Omega^{2}},      
\end{equation}
with $\mathcal{M}(r)$ a smooth function for $r > 0$ (i.e. $\mathcal{M}(r)\in\mathcal{C}^{^{\infty}}$). 
Notice that this is the most general form of the metric consistent with our assumptions $g_{tt}g_{rr}=-1$, the equations of motion, and the chosen coordinate system.

The event horizon is located at $r=r_{h}$ provided that
%%%%%%%%%%
\begin{equation}
\mathcal{M}(r_{h}) = \frac{r_{h}}{2}, \quad\quad\quad\quad \mathcal{M}(r) < \frac{r}{2} \quad\quad \forall r > r_{h}, \quad\quad\quad \textup{and}\quad\quad\quad  \mathcal{M}(r \rightarrow \infty) = \textup{constant} \geq0.  
\end{equation}
The new geometry of the SSS-AF regular black hole type will be denoted by 
$\boldsymbol{d\tilde s^{2}}= \tilde g_{\alpha\beta}\boldsymbol{dx^{\alpha}dx^{\beta}}$.
In particular, we shall adopt the form $\boldsymbol{d\tilde s^{2}}
= \tilde g_{tt}\boldsymbol{dt^{2}} +  \tilde g_{rr}
\boldsymbol{dr^{2}} + \tilde g_{\theta\theta}
\boldsymbol{d\theta^{2}} + \tilde g_{\varphi\varphi} 
\boldsymbol{d\varphi^{2}}$, 
with
\begin{eqnarray}
&&\tilde g_{tt}
= -1 + \left(1 - \frac{a^{2}}{r^{2}}\right)\!\!\left(1 + g_{tt}
\right), \quad\quad   \tilde g_{rr}
= \left(1 - \frac{a^{2}}{r^{2}}\right)^{\!\!\!\!^{-1}} \left[ 1 - \left(1 - \frac{a^{2}}{r^{2}}\right)\!\!\left(1 - \frac{1}{g_{rr}
}\right) \right]^{\!\!^{-1}}, \\
\nonumber\\
&&\tilde g_{\theta\theta}
= \left(1 - \frac{a^{2}}{r^{2}}\right)g_{\theta\theta}
, \quad\quad\quad\quad\quad\quad\quad \tilde g_{\varphi\varphi}
= \left(1 - \frac{a^{2}}{r^{2}}\right)g_{\varphi\varphi},
\end{eqnarray}
where $a$ is a real parameter to be related later to physical quantities.
Using Eq.\eqref{BH_structure}, it follows that
\begin{equation}\label{RBH_structure}
\boldsymbol{d\tilde s^{2}}
= - \left[ 1 - \frac{2 \mathcal{M}(r)}{r}\!\!\left( 1 - \frac{a^{2}}{r^{2}} \right) \right]\boldsymbol{dt^{2}} + \frac{\boldsymbol{dr^{2}}}{\left( 1 - \frac{a^{2}}{r^{2}} \right)\!\left[ 1 - \frac{2\mathcal{M}(r)}{r}\!\!\left( 1 - \frac{a^{2}}{r^{2}}\right) \right] }
+ (r^{2}-a^{2})\boldsymbol{d\Omega^{2}}.     
\end{equation}
This metric admits a Lorentzian signature in the region $r>|a|$, while in the region $r\in(0,|a|)$ the metric signature could be either $(----)$ or $(++--)$, depending of the behavior of $\mathcal{M}(r)$.
In Appendix \ref{AppendixB}, we present
the expression of  some curvature invariants for the metric in Eq.\eqref{RBH_structure}, which are finite. Since all the curvature invariants behave as those in Appendix \ref{AppendixB}, the geometry is regular over the part of the manifold covered by the metric in Eq.(\ref{RBH_structure}), \emph{i.e.}  for $r>|a|$.  \\
%%%%%%%%
By means of the transformation
$\rho^{2} =  r^{2} - a^{2} $, with $\boldsymbol{d\rho} = \pm (r/\sqrt{r^{2} - a^{2}})\boldsymbol{dr}$, Eq.\eqref{RBH_structure} can be written as
%%%%%%% 
\begin{equation}\label{RBH_structure_SV}
\boldsymbol{d\tilde s^{2}}
= - \left( 1 - \frac{2\hskip.04cm\mathcal{M}\hskip.04cm\rho^{2} }{ \left(\rho^{2} + a^{2}\right)^{\frac{3}{2}} } \right)\boldsymbol{dt^{2}} + \left( 1 - \frac{2\hskip.04cm\mathcal{M}\hskip.04cm\rho^{2}}{ \left(\rho^{2} + a^{2}\right)^{\frac{3}{2}} } \right)^{-1}
\boldsymbol{d\rho^{2}}
+ \rho^{2}\boldsymbol{d\Omega^{2}} \quad\quad\textup{with}\quad\quad \mathcal{M}=\mathcal{M}(r(\rho^{2})),
\end{equation}
in which the Schwarzschild-like coordinates ($t,\rho,\theta,\varphi$) take values as follows:
$t\in(-\infty,\infty)$, $\rho\in(0,\infty)$, 
$\theta\in[0,\pi]$ and $\varphi\in[0,2\pi)$. In contrast to the metric in Eq.(\ref{RBH_structure}), the unphysical region $r<|a|$ is not covered by the metric in Eq.(\ref{RBH_structure_SV}).\\
%%%%%%%%%%%%%%%%%%%%%%%%%%%%%%%
%%%%%%%%%%%%%%%%%%%%%%%%%%%%%%%%%%%%%%%%%%%%%%%%%%%%%%%%%%%%%%%%%%%%%%%%%

\section{Static spherically symmetric regular black hole geometries in general relativity coupled to non-linear electrodynamics}\label{BBinGR-NLED}

The GR-NLED action is given by
%%%%%%%%%%%%%%%%%%%%%%%%%%%%%%%%%%%%
\begin{equation}\label{actionL}
S[g_{ab},A_{a}] = \frac{1}{16\pi} \int (R - 4\mathcal{L}(\mathcal{F}))\quad\!\!\!\! \sqrt{-g} \quad\!\!\!\! {d^{4}x},
\end{equation}
where $R$ is the scalar curvature, and $\mathcal{L}$ is a function of the electromagnetic invariant $\mathcal{F}\equiv \frac{1}{4}F_{\alpha\beta}F^{\alpha\beta}$, with $F_{\alpha\beta}=2\partial_{[\alpha}A_{\beta]}$  
the components of the electromagnetic field tensor  $\boldsymbol{F}=\frac{1}{2}F_{\alpha\beta} \boldsymbol{dx^{\alpha}} \wedge \boldsymbol{dx^{\beta}}$, and $A_{a}$, the components of the electromagnetic potential. 

The GR-NLED field equations obtained from the variation of Eq.(\ref{actionL}) are, 
\begin{equation}\label{modifEqs} 
G_{\alpha}{}^{\beta} \!=\!8\pi (E_{\alpha}{}^{\beta})\!_{_{_{N \! L \! E \! D}}}, 
\end{equation}
\begin{equation}
\label{e1}
\nabla_{\mu}(\mathcal{L}_{\mathcal{F}}F^{\mu\nu})\!=\!0,
\end{equation}
with $
4\pi(E_{\alpha}{}^{\beta})\!_{_{_{N \! L \! E \! D}}} = \mathcal{L}_{\mathcal{F}} \hskip.01cm F_{\alpha\mu}F^{\beta\mu} - \mathcal{L}\hskip.01cm\delta_{\alpha}{}^{\beta}$,
where 
$(E_{\alpha}{}^{\beta})\!_{_{_{N \! L \! E \! D}}}$  denotes
the components of the NLED energy-momentum tensor,
$\mathcal{L}_{\mathcal{F}}\equiv \frac{d\mathcal{L}}{d\mathcal{F}}, $
and $G_{\alpha}{}^{\beta} = R_{\alpha}{}^{\beta} - \frac{R}{2} \delta_{\alpha}{}^{\beta}$.
The identity 
\begin{equation}
\label{e2}
\nabla_{\![\lambda} \hspace{0.03cm} F_{\mu\nu]}=0  
\end{equation}
must also be considered when solving for the fields of the system.
%
%%%%%%%%%%%%%%%%%%

An alternative representation for GR-NLED
can be defined using the auxiliary anti-symmetric tensor $P_{\mu\nu} = \mathcal{L}_{\mathcal{F}} F_{\mu\nu}$, and 
the function $\mathcal{H}$, obtained from 
the Lagrangian density
by means of a Legendre transformation
(see for instance \cite{Gutierrez1981}),
\begin{equation}\label{actionH}
\mathcal{H} = 2\mathcal{F}\mathcal{L}_{\mathcal{F}} -  \mathcal{L}. 
\end{equation} 
It can be shown that $\mathcal{H}$ is a function of the 
invariant $\mathcal{P}$, defined as $\mathcal{P} = \frac{1}{4}P_{\mu\nu}P^{\mu\nu} = (\mathcal{L}_{\mathcal{F}})^{2}\mathcal{F}$. 

Regarding the electromagnetic field tensor, since the spacetime is static and spherically symmetric, the only
non-vanishing terms are the electric $F_{tr}$ and magnetic $F_{\theta\phi}$ components, in such a way that
%%%%
\begin{equation}\label{Fab_SSS}
F_{\alpha\beta} = \Big( \delta^{t}_{\alpha}\delta^{r}_{\beta} - \delta^{r}_{\alpha}\delta^{t}_{\beta}\Big) F_{tr} + \Big( \delta^{\theta}_{\alpha}\delta^{\varphi}_{\beta} - \delta^{\varphi}_{\alpha}\delta^{\theta}_{\beta}\Big) F_{\theta\varphi}.       
\end{equation}
Hence, the non-vanishing components of the energy-momentum tensor for NLED, obtained from  
the metric given in Eq.(\ref{RBH_structure}), with the electromagnetic field tensor in Eq.(\ref{Fab_SSS}), and a Lagrangian density $\mathcal{L}(\mathcal{F})$, are given by,
\begin{eqnarray}
&& 8\pi (E_{t}{}^{t})\!_{_{_{N \! L \! E \! D}}} = 8\pi (E_{r}{}^{r})\!_{_{_{N \! L \! E \! D}}} = 2(F_{tr}F^{tr}\mathcal{L}_{\mathcal{F}}-\mathcal{L}), \label{NLEDtt}\\
&&8\pi (E_{\theta}{}^{\theta})\!_{_{_{N \! L \! E \! D}}} = 8\pi (E_{\varphi}{}^{\varphi})\!_{_{_{N \! L \! E \! D}}} = 2(F_{\theta\varphi}F^{\theta\varphi}\mathcal{L}_{\mathcal{F}}-\mathcal{L}). 
\end{eqnarray}
Using Eq.\eqref{e1}, written as $\partial_{\alpha}(\sqrt{-g}\mathcal{L}_{\mathcal{F}}F^{\alpha\beta}) = 0$, and Eq.\eqref{e2}, it follows that
%
%%%%%
\begin{equation}
F^{rt} = \frac{ 
q_{_{\mathcal{E}}} }{(r^{2}-a^{2})\sqrt{-g_{tt}g_{rr}}\mathcal{L}_{\mathcal{F}} }, \quad\quad\quad\quad\quad
F^{\theta\varphi} = \frac{ q_{_{\mathcal{B}}}}{(r^2-a^2)^2 \sin\theta },
\end{equation}
where $q_{_{\mathcal{E}}}$ and $q_{_{\mathcal{B}}}$
are the electric and magnetic charges, respectively.
Thus, the non-null
components of $F_{\alpha\beta}$  and the invariant $\mathcal{F}$ for the spacetime geometry in Eq.(\ref{RBH_structure}) are given by
%%%%%%%
\begin{equation}\label{F_invariant}
F_{tr} = \frac{ q_{_{\mathcal{E}}} r }{(r^{2}-a^{2})^{^{\frac{3}{2}}}\mathcal{L}_{\mathcal{F}}},\quad\quad\quad  F_{\theta\varphi} =   q_{_{\mathcal{B}}} \sin\theta, \quad\quad\quad \mathcal{F} =  \frac{1}{2(r^{2}-a^{2})^{2}}\left(q^{2}_{_{\mathcal{B}}}  - \frac{ q^{2}_{_{\mathcal{E}}} }{\mathcal{L}^{^{2}}_{\mathcal{F}}}\right)
\end{equation}
%%%%%%%
with $a$, $q_{_{\mathcal{E}}}$ and $q_{_{\mathcal{B}}}$ constants\footnote{ It follows from the decomposition of $F_{\mu\nu}$ in terms of $E_\mu$, $H_\mu$ (being $(E,H)$ the electric and magnetic field strengths respectively), and $v_\mu$ (the velocity of the observer, see for instance 
\cite{Novello2002}),
that $F_{\theta\varphi}$ as given in Eq.\eqref{F_invariant} leads to a spherically symmetric $H^\mu$, corresponding to a point magnetic charge. }. 
We shall adopt the relation
$a^{2} = \sigma_{_{\mathcal{E}}} \quad\!\!\!\!\! q_{_{\mathcal{E}}}^{2}  + \sigma_{_{\mathcal{B}}} \quad\!\!\!\!\! q_{_{\mathcal{B}}}^{2}$ under which, as will be shown below, all the cases presented in the literature can be reproduced with the procedure displayed in the previous section. Such a relation is consistent with the idea that 
the removal of the singularity of the original BH solution is solely due to NLED, with
$\sigma_{_{\mathcal{E}}}$ and $\sigma_{_{\mathcal{E}}}$ positive parameters such that  $\sigma_{_{\mathcal{E}}}=\sigma_{_{\mathcal{B}}}=0$ if $\mathcal{L}_{\mathcal{F}} = $ constant for all $\mathcal{F}$ (\emph{i.e.} linear electrodynamics), and  $\sigma_{_{\mathcal{E}}}\neq0\neq\sigma_{_{\mathcal{B}}}$ if $\mathcal{L}_{\mathcal{F}} \neq$ constant (\emph{i.e.} non-linear electrodynamics).  

The non-null components of $P_{\alpha\beta}$ and the invariant $\mathcal{P}$ are respectively given by
\begin{equation}\label{P_invariant}
P_{tr} = \frac{ q_{_{\mathcal{E}}} r }{(r^{2}-a^{2})^{^{\frac{3}{2}}}},\quad\quad\quad  P_{\theta\varphi} = \frac{ q_{_{\mathcal{B}}} \sin\theta}{\mathcal{H}_{\mathcal{P}}}, \quad\quad\quad
\mathcal{P} = \frac{1}{2(r^{2}-a^{2})^{2}}\left(\frac{q^{2}_{_{\mathcal{B}}}}{\mathcal{H}^{^{2}}_{\mathcal{P}}}  - q^{2}_{_{\mathcal{E}}}  
\right),
\end{equation}
%%%%
where
$\mathcal{H}_{\mathcal{P}}\equiv \frac{d\mathcal{H}}{d\mathcal{P}} $.
Thus, Eq.(\ref{F_invariant}) is the general solution of Eqs.\eqref{e1} and \eqref{e2} for static and spherically symmetric electromagnetic fields in the $\mathcal{L}(\mathcal{F})$ representation, while Eq.(\ref{P_invariant}) is the general solution in the $\mathcal{H}(\mathcal{P})$ representation. 

Next we shall use these expressions to solve the GR-NLED field equations, ${C}_{\alpha}{}^{\beta} \equiv G_{\alpha}{}^{\beta} - 8\pi(E_{\alpha}{}^{\beta})\!_{_{_{N \! L \! E \! D}}} = 0$ in several cases of interest. In particular, we shall see that the configurations obtained with the procedure presented in the previous section are in accordance with the theorem presented in \cite{Bronnikov2000}, which states that
``the field equations following from Eq.\eqref{actionL},
with ${\cal L}({\cal F})$ having a
Maxwell asymptotic limit (${\cal L}\rightarrow  0, {\cal L}_{\cal F} \rightarrow 1$ as ${\cal F}\rightarrow  0$), do
not admit a static, spherically symmetric solution with a
regular center and a nonzero electric charge '' (thus including the dyonic case). 

\subsection{Purely Magnetic Solutions}

The relevant GR-NLED field equations for the purely magnetic case ($F_{tr}=q_{_{\mathcal{E}}}=0$, $F_{\theta\varphi} \neq0\neq q_{_{\mathcal{B}}}=q$) in the $\mathcal{L}(\mathcal{F})$ representation
are more conveniently obtained
from Eqns.\eqref{eq1} and \eqref{eq2}, and are given by
\begin{eqnarray}
&&\frac{2(a^{2} - r^{2})\mathcal{M}'}{ r^{4}} - \frac{6a^{2}\mathcal{M}}{ r^{5}} = - 2\mathcal{L},\label{MagGR_NLEDrt}\\
&&-\frac{\left( r^{2} - a^{2}\right)^{2}\mathcal{M}''}{ r^{5}} + \frac{7a^{2}\left( a^{2} - r^{2}\right)\mathcal{M}'}{ r^{6}} + \frac{3a^{2}\left( 3r^{2} - 5a^{2}\right)\mathcal{M}}{ r^{7}}
= \frac{2q^{2}\mathcal{L}_{\mathcal{F}}}{(r^{2}-a^{2})^{^{2}}} 
- 2\mathcal{L}, \label{MagGR_NLEDaa}
\end{eqnarray}
with $a^{2} = \sigma_{_{\mathcal{B}}} \quad\!\!\!\!\! q^{2}$. 
For a given 
$\mathcal{M}(r)$, the electromagnetic Lagrangian $\mathcal{L}$ as function of the $r$-coordinate can be obtained from Eq.\eqref{MagGR_NLEDrt}, and $\mathcal{L}=\mathcal{L}(\mathcal{F})$ 
follows using $r=r(\mathcal{F})$, by way of inverting Eq.\eqref{F_invariant} with $q_{_{\mathcal{E}}}=0$. \\
The case 
$q=0$ ({\it{i.e.}} null magnetic field)
leads to $\mathcal{L}=0$ and  $a^{2}=0$. Hence, the general solution of the field equations (\ref{MagGR_NLEDrt})-(\ref{MagGR_NLEDaa}) is $\mathcal{M}(r)=m=$ constant $\in\mathbb{R}$. Therefore, in this case the line element (\ref{RBH_structure}) $\equiv$ (\ref{RBH_structure_SV}) becomes the Schwarzschild metric. 
%%%%%%%%%%%%%%%%%%%%%%%%%%%%%%%%%%%%%%%%%%%%%%%%%%%%%%%%%%%%%%%%%%%%%%%%%%%%%%%%%%%%%%%%%%%%%%%%%%%%%%%%%%%%%%%%%
%%%%%%%%%%%%%%%%%%%%%%%%%%%%%%%%%%%%%%%%%%%%%%%%%%%%%%%%%%%%%%%%%%%%%%%%%%%%%%%%%%%%%%%%%%%%%%%%%%%%%%%%%%%%%%%%%%%
%%%%%%%%%%%%%%%%%%%%%%%%%%%%%%%%%%%%%%%%%%%%%%%%%%%%%%%%%%%%%%%%%%%%%%%%%%%%%%%%%%%%%%%%%%%%%%%%%%%%%%%%%%%%%%%%%%%
%%%%%%%%%%%%%%%%%%%%%%%%%%%%%%%%%%%%%%%%%%%%%%%%%%%%%%%%%%%%%%%%%%%%%%%%%%%%%%%%%%%%%%%%%%%%%%%%%%%%

\subsubsection{ {\bf Applications } }

\begin{itemize}

\item {\it Removal of the singularity of the Schwarzschild black hole:} For the case $-g_{_{tt}}
= 1/g_{_{rr}} = 1-2m/r$, with $m$ = constant $\in\mathbb{R}^{+}$, the line element in Eq.(\ref{RBH_structure}) takes the form  
\begin{equation}\label{RBH_Schw_structure}
\boldsymbol{d\tilde s^{2}} = - \left[ 1 - \frac{2m}{r}\!\!\left( 1 - \frac{\sigma_{_{\mathcal{B}}}\quad\!\!\!\!\! q^{2}}{r^{2}} \right) \right]\boldsymbol{dt^{2}} + \frac{\boldsymbol{dr^{2}}}{\left( 1 - \frac{\sigma_{_{\mathcal{B}}}\quad\!\!\!\!\!\!\! q^{2}}{r^{2}} \right)\!\left[ 1 - \frac{2m}{r}\!\!\left( 1 - \frac{ \sigma_{_{\mathcal{B}}}\quad\!\!\!\!\!\!\! q^{2} }{r^{2}}\right) \right] }
+ \left(r^{2}-\sigma_{_{\mathcal{B}}}\quad\!\!\!\!\! q^{2}\right)\boldsymbol{d\Omega^{2}}.     
\end{equation}
The corresponding NLED model for which this line element is an exact purely magnetic solution of the field equations (\ref{MagGR_NLEDrt})-(\ref{MagGR_NLEDaa}) is described by the Lagrangian 
\begin{equation}
\mathcal{L}(\mathcal{F}) 
= \frac{ 3 \quad\!\!\!\!\! \sigma_{_{\mathcal{B}}} }{2\quad\!\!\!\!\! s \quad\!\!\!\!\! q^{2} }\left( \frac{ \sqrt{2 \quad\!\!\!\!\! q^{2} \quad\!\!\!\!\! \mathcal{F}} }{ 1 + \sqrt{ 2\quad\!\!\!\!\! \sigma^{2}_{_{\mathcal{B}}}\quad\!\!\!\!\! q^{2} \mathcal{F} }  } \right)^{\frac{5}{2}} \quad\quad\textup{ or }\quad\quad \mathcal{L}(r) = \frac{3\quad\!\!\!\!\! \sigma_{_{\mathcal{B}}}\quad\!\!\!\!\!m\quad\!\!\!\!\! q^{2}  }{r^{5}},    
\end{equation}
where 
$s=|q|/(2m)$; and $m$ is a free parameter associated with the mass of the configuration. 

Note that, for the case under consideration, $\mathcal{M}=m=$ constant $\in\mathbb{R}^{+}$, and the inequalities
(\ref{WEC_1})
that define the WEC and the NEC are satisfied.  
Finally, the application of the mapping $\rho^{2} =  r^{2} - \sigma_{_{\mathcal{B}}}\quad\!\!\!\!\!q^{2}$; $\boldsymbol{d\rho} = \pm (r/\sqrt{r^{2} - \sigma_{_{\mathcal{B}}}\quad\!\!\!\!\! q^{2}})\boldsymbol{dr}$ to Eq.\eqref{RBH_Schw_structure}
yields
\begin{equation}\label{RBH_structure_SVs}
\boldsymbol{d\tilde s^{2}} = - \left( 1 - \frac{2 m \rho^{2} }{ \left(\rho^{2} + \sigma_{_{\mathcal{B}}} \quad\!\!\!\!\! q^{2}\right)^{\frac{3}{2}} } \right)\boldsymbol{dt^{2}} + \left( 1 - \frac{2 m \rho^{2}}{ \left(\rho^{2} + \sigma_{_{\mathcal{B}}} \quad\!\!\!\!\! q^{2}\right)^{\frac{3}{2}} } \right)^{-1}
\boldsymbol{d\rho^{2}}
+ \rho^{2}\boldsymbol{d\Omega^{2}} 
\end{equation}
with
$\rho\in[0,\infty)$. 
This metric reduces to that of the 
Bardeen black hole \cite{Bardeen1968} for 
$\sigma_{_{\mathcal{B}}}=1$ (and  
the function (\ref{L_NLED}) becomes that of the Bardeen model). For $\sigma_{_{\mathcal{B}}}=0$ the line element (\ref{RBH_structure_SVs}) becomes that of the  Schwarzschild BH, and the electromagnetic effects are turned off (i.e. $\mathcal{L}(\mathcal{F}) = 0$, from Eq.(\ref{L_NLED})) .

This first example shows how a well-known solution of the GR-NLED system can actually be obtained using our procedure from a singular solution of GR.

\item {\it Removal of the singularity of the canonical acoustic black hole:} 
For the case $-g_{_{tt}}= 1/g_{_{rr}}= 1-\mu/r^{4}$, being $\mu$ a positive real constant,  
which corresponds to a spherically symmetric
flow of an incompressible fluid
(see for details \cite{Visser1998,Canate2021}), the line element (\ref{RBH_structure}) takes the form  %%%
%%%
\begin{equation}\label{RBH_CABH_structure}
\boldsymbol{d\tilde s^{2}} = - \left[ 1 - \frac{\mu}{r^{4}}\!\!\left( 1 - \frac{\sigma_{_{\mathcal{B}}}\quad\!\!\!\!\! q^{2}}{r^{2}} \right) \right]\boldsymbol{dt^{2}} + \frac{\boldsymbol{dr^{2}}}{\left( 1 - \frac{\sigma_{_{\mathcal{B}}}\quad\!\!\!\!\!\!\! q^{2}}{r^{2}} \right)\!\left[ 1 - \frac{\mu}{r^{4}}\!\!\left( 1 - \frac{ \sigma_{_{\mathcal{B}}}\quad\!\!\!\!\!\!\! q^{2} }{r^{2}}\right) \right] } + \left(r^{2}-\sigma_{_{\mathcal{B}}}\quad\!\!\!\!\! q^{2}\right)\boldsymbol{d\Omega^{2}}. \end{equation}

The NLED Lagrangian density corresponding to this line element, for which the source is purely magnetic, is obtained from the field equations (\ref{MagGR_NLEDrt})-(\ref{MagGR_NLEDaa}), and given by
%%%%%%
\begin{equation}
\mathcal{L}(\mathcal{F}) = \frac{ 3 }{4\quad\!\!\!\!\! s \quad\!\!\!\!\! |q|^{5} } \frac{  \left(  \sqrt{ 2\quad\!\!\!\!\! \sigma^{2}_{_{\mathcal{B}}}\quad\!\!\!\!\! q^{2} \mathcal{F} } - 1 \right) \left(2 q^{2} \mathcal{F} \right)^{\frac{3}{2}}  }{ \left( 1 + \sqrt{ 2\quad\!\!\!\!\! \sigma^{2}_{_{\mathcal{B}}}\quad\!\!\!\!\! q^{2} \mathcal{F} } \right)^{4} }  \quad\quad\textup{ or }\quad\quad \mathcal{L}(r) = \frac{3\mu(2\quad\!\!\!\!\! \sigma_{_{\mathcal{B}}}\quad\!\!\!\!\!q^{2} - r^{2} )}{2r^{8}},
\end{equation}
with $\mu=|q|/(2s)$. 
In the weak field limit,  this Lagrangian goes to a power law, a case  to the which was studied for instance in 
\cite{Gonzalez2009}.

The mapping $\rho^{2} = r^{2} - \sigma_{_{\mathcal{B}}}\quad\!\!\!\!\! q^{2}$; $\boldsymbol{d\rho} = \pm (r/\sqrt{r^{2} - \sigma_{_{\mathcal{B}}}\quad\!\!\!\!\! q^{2}})\boldsymbol{dr}$ takes the metric given in Eq.\eqref{RBH_CABH_structure} to the form
%%%%%%%%%%%%%%%%%%%%%%%%
\begin{equation}\label{RBH_structure_c1}
\boldsymbol{d\tilde s^{2}}
= - \left( 1 - \frac{ \mu \quad\!\!\!\!\! \rho^{2}}{\left(\rho^{2} + \sigma_{_{\mathcal{B}}} \quad\!\!\!\!\! q^{2}\right)^{3} } \right)\boldsymbol{dt^{2}} + \left( 1 - \frac{\mu \quad\!\!\!\!\! \rho^{2}}{ \left(\rho^{2} + \sigma_{_{\mathcal{B}}} \quad\!\!\!\!\! q^{2}\right)^{3} } \right)^{-1}
\boldsymbol{d\rho^{2}}
+ \rho^{2}\boldsymbol{d\Omega^{2}}
\end{equation}
 with $\rho\in[0,\infty)$. 
Hence, in this example the procedure was applied to a black hole spacetime which is not a vacuum, or electro-vacuum, 
solution of the GR equations, and transformed into a regular BH which is solution of the system GR-NLED.
Notice also that the same ``seed metric'' (namely, that of the canonical acoustic black hole) could be used to generate a purely electric solution (see next section).

\end{itemize}

%%%%%%%%%%%%%%%%%%%%%%%%%%%%%%%%%%%%%%%%%%%%%%%%%%%%%%%%%%%%%%%%%%%%%%%%%%%%%%%%%%%%%%%%%%%%%%%%%%%%%%%%%%%%%%%%%%%
%%%%%%%%%%%%%%%%%%%%%%%%%%%%%%%%%%%%%%%%%%%%%%%%%%%%%%%%%%%%%%%%%%%%%%%%%%%%%%%%%%%%%%%%%%%%%%%%%%%%%%%%%%%%%%%%%%%
%%%%%%%%%%%%%%%%%%%%%%%%%%%%%%%%%%%%%%%%%%%%%%%%%%%%%%%%%%%%%%%%%%%%%%%%%%%%%%%%%%%%%%%%%%%%%%%%%%%%%%%%%%%%%%%%%%%
%%%%%%%%%%%%%%%%%%%%%%%%%%%%%%%%%%%%%%%%%%%%%%%%%%%%%%%%%%%%%%%%%%%%%%%%%%%%%%%%%%%%%%%%%%%%%%%%%%%%%%%%%%%%%%%%%%%

\subsection{Purely Electric Solutions}

The GR-NLED field equations for the purely electric case ($F_{\theta\varphi}=q_{_{\mathcal{B}}}=0$, $F_{tr} \neq0\neq q_{_{\mathcal{E}}}=q$) in the $\mathcal{H}(\mathcal{P})$
representation can be directly obtained from Eqns.\eqref{eq3} and \eqref{eq4}:
\begin{eqnarray}
&&\frac{2(a^{2} - r^{2})\mathcal{M}'}{ r^{4}} - \frac{6a^{2}\mathcal{M}}{ r^{5}} =  2\mathcal{H}, \label{ElecGR_NLEDrt} \\
%%%%%
&&-\frac{\left( r^{2} - a^{2}\right)^{2}\mathcal{M}''}{ r^{5}} + \frac{7a^{2}\left( a^{2} - r^{2}\right)\mathcal{M}'}{ r^{6}} + \frac{3a^{2}\left( 3r^{2} - 5a^{2}\right)\mathcal{M}}{ r^{7}}=\frac{2q^{2}\mathcal{H}_{\mathcal{P}}}{(r^{2}-a^{2})^{^{2}}} + 2\mathcal{H}, \label{ElecGR_NLEDaa}
\end{eqnarray} %% 
with $a^{2} = \sigma_{_{\mathcal{E}}} \quad\!\!\!\!\! q^{2}$.
%%%%%%%%%
%%
For a given 
$\mathcal{M}(r)$ the characteristic function $\mathcal{H}$ as function of the $r$-coordinate follows from Eq.\eqref{ElecGR_NLEDrt}, and then $\mathcal{H}=\mathcal{H}(\mathcal{P})$ 
can be obtained using $r=r(\mathcal{P})$ by way of inverting Eq.\eqref{P_invariant} with $q_{_{\mathcal{B}}}=0$. \\
For zero electric field ($q=0$), we have that 
$\mathcal{H}=0$, and  $a^{2}=0$. Hence, the general solution of Eqns.(\ref{ElecGR_NLEDrt})-(\ref{ElecGR_NLEDaa}) is $\mathcal{M}(r)=m=$ constant $\in\mathbb{R}$. Therefore for this case the line element (\ref{RBH_structure}) $\equiv$ (\ref{RBH_structure_SV}) becomes the Schwarzschild metric.

\subsubsection*{ {\bf Applications } }

\begin{itemize}

\item {\it Removal of the singularity of the Reissner-Nordstr\"{o}m black hole:}  $-g_{_{tt}} = 1/g_{_{rr}} = 1 - 2m/r + q^{2}/r^{2}$, with $m$, $q$ real constants.
%%%%%
In this case, the line element in Eq.(\ref{RBH_structure}) takes the form
\begin{equation}\label{RBH_ReissN_structure}
\boldsymbol{d\tilde s^{2}} = - \left[ 1 - \left(\frac{2m}{r}-\frac{q^{2}}{r^{2}}\right)\!\!\left( 1 - \frac{\sigma_{_{\mathcal{E}}} \quad\!\!\!\!\!q^{2}}{r^{2}} \right) \right]\boldsymbol{dt^{2}} + \frac{\boldsymbol{dr^{2}}}{\Big( 1 - \frac{\sigma_{_{\mathcal{E}}} \quad\!\!\!\!\!\!\!q^{2}}{r^{2}} \Big)\!\left[ 1 - \left(\frac{2m}{r}-\frac{q^{2}}{r^{2}}\right)\!\!\Big( 1 - \frac{\sigma_{_{\mathcal{E}}} \quad\!\!\!\!\!\!\!q^{2}}{r^{2}}\Big) \right] }
+ \left(r^{2} - \quad \!\!\!\!\!\sigma_{_{\mathcal{E}}} \quad\!\!\!\!\! q^{2}\right)\boldsymbol{d\Omega^{2}}.     
\end{equation}
Notice that in this case an electric charge, $q$, is already present in the seed metric, and extra charge comes from the application of the method described in Sect. II. The NLED model for which this line element is an exact pure electric solution of the field equations (\ref{ElecGR_NLEDrt})-(\ref{ElecGR_NLEDaa}) is given by
%%%%%
\begin{equation}
\label{hag}
\mathcal{H}(\mathcal{P}) =  \frac{ \left( 1 - 3 \sqrt{ -2 \quad\!\!\!\!\!\sigma^{2}_{_{\mathcal{E}}} \quad\!\!\!\!\! q^{2} \quad\!\!\!\!\! \mathcal{P} } \right) \mathcal{P} }{ \left( 1 + \sqrt{ -2\quad\!\!\!\!\! \sigma^{2}_{_{\mathcal{E}}} \quad\!\!\!\!\! q^{2} \quad\!\!\!\!\!\mathcal{P} } \right)^{3} } - \frac{ 3 \quad\!\!\!\!\! \sigma_{_{\mathcal{E}}} }{2 s q^{2}}\left( \frac{ \sqrt{-2q^{2}\mathcal{P}} }{ 1 + \sqrt{ -2\quad\!\!\!\!\! \sigma^{2}_{_{\mathcal{E}}}\quad\!\!\!\!\! q^{2}\mathcal{P} }  } \right)^{\frac{5}{2}} %%%% %%%%
\quad\textup{ or }\quad\mathcal{H}(r) = \frac{q^{2}(4\sigma_{_{\mathcal{E}}} q^{2} - r^{2})}{2r^{6}} - \frac{3 \sigma_{_{\mathcal{E}}} q^{2}m  }{r^{5}},
\end{equation}
where $s$ stands for $s=|q|/(2m)$, and $m\geq0$ is the mass of the configuration.\\
Finally, the mapping $\rho^{2} =  r^{2} - \sigma_{_{\mathcal{E}}} \quad\!\!\!\!\! q^{2}$; $\boldsymbol{d\rho} = \pm (r/\sqrt{r^{2} - \sigma_{_{\mathcal{E}}} \quad\!\!\!\!\! q^{2}})\boldsymbol{dr}$ yields
\begin{equation}\label{RBH_structure_ReissN}
\boldsymbol{d\tilde s^{2}}
= - \left( 1 - \frac{2 m \rho^{2} }{ \left(\rho^{2} +  \sigma_{_{\mathcal{E}}} \quad\!\!\!\!\! q^{2}\right)^{\frac{3}{2}} } + \frac{ q^{2} \rho^{2} }{ \left(\rho^{2} +  \sigma_{_{\mathcal{E}}} \quad\!\!\!\!\! q^{2}\right)^{2} } \right)\boldsymbol{dt^{2}} + \left( 1 - \frac{2 m \rho^{2}}{ \left(\rho^{2} + \sigma_{_{\mathcal{E}}} \quad\!\!\!\!\! q^{2}\right)^{\frac{3}{2}} } + \frac{ q^{2} \rho^{2} }{ \left(\rho^{2} + \sigma_{_{\mathcal{E}}} \quad\!\!\!\!\! q^{2}\right)^{2} }\right)^{-1}
\boldsymbol{d\rho^{2}}
+ \rho^{2}\boldsymbol{d\Omega^{2}}. 
\end{equation}
The  Ay\'on-Garc\'ia black hole solution
and the corresponding
$\mathcal{H}$
function 
\cite{Ayon1998} are recovered
in the case $\sigma_\mathcal{E} = 1$.
It is important to point out that, in contrast to the Ay\'on-Garc\'ia black hole solution which does not have the Reissner-Nordstr\"{o}m solution as a particular case, our solution (\ref{RBH_structure_ReissN}) 
becomes 
the Reissner-Nordstr\"{o}m solution if $\sigma_{_{\mathcal{E}}}=0$, and Eq.(\ref{H_NLED}) becomes the linear model $\mathcal{H}(\mathcal{P}) = \mathcal{P}$.  
Notice that in the weak-field limit, Eq.\eqref{hag} yields $\mathcal{H}\approx  \mathcal{P}$, a fact that  would be in accordance with a theorem dual to the one mentioned above, stated in the $\mathcal{H}$ frame.
%%%%%%%%%%%%%%%%%%%%%%%%%%%%%%%%%%%%%%%%%%%%%%%%%%

\item {\it Removal of the singularity of the canonical acoustic black hole using a pure electric field:}  
For the case $-g_{_{tt}}= 1/g_{_{rr}}= 1-\mu/r^{4}$, with $\mu=$ constant $\in\mathbb{R}^{+}$,  
the line element (\ref{RBH_structure}) with $a^{2} = \sigma_{_{\mathcal{E}}} \quad\!\!\!\!\! q_{_{\mathcal{E}}}^{2}$ becomes,  
%%%
\begin{equation}\label{RBH_CABH_structure_mag}
\boldsymbol{d\tilde s^{2}} = - \left[ 1 - \frac{\mu}{r^{4}}\!\!\left( 1 - \frac{\sigma_{_{\mathcal{E}}}\quad\!\!\!\!\! q^{2}}{r^{2}} \right) \right]\boldsymbol{dt^{2}} + \frac{\boldsymbol{dr^{2}}}{\left( 1 - \frac{\sigma_{_{\mathcal{E}}}\quad\!\!\!\!\!\!\! q^{2}}{r^{2}} \right)\!\left[ 1 - \frac{\mu}{r^{4}}\!\!\left( 1 - \frac{ \sigma_{_{\mathcal{E}}}\quad\!\!\!\!\!\!\! q^{2} }{r^{2}}\right) \right] } + \left(r^{2}-\sigma_{_{\mathcal{E}}}\quad\!\!\!\!\! q^{2}\right)\boldsymbol{d\Omega^{2}}. \end{equation}
The  corresponding to this line element, for which the source is purely electric, is obtained from the field equations (\ref{ElecGR_NLEDrt})-(\ref{ElecGR_NLEDaa}), and given by
%%%%%%
\begin{equation}
\mathcal{H}(\mathcal{P}) = \frac{ 3 }{4\quad\!\!\!\!\! s \quad\!\!\!\!\! |q|^{5} } \frac{  \left( 1 -  \sqrt{ -2\quad\!\!\!\!\! \sigma^{2}_{_{\mathcal{E}}}\quad\!\!\!\!\! q^{2} \mathcal{P} }  \right) \left(-2 q^{2} \mathcal{P} \right)^{\frac{3}{2}}  }{ \left( 1 + \sqrt{- 2\quad\!\!\!\!\! \sigma^{2}_{_{\mathcal{E}}}\quad\!\!\!\!\! q^{2} \mathcal{P} } \right)^{4} }  \quad\quad\textup{ or }\quad\quad \mathcal{H}(r) = -\frac{3\mu(2\quad\!\!\!\!\! \sigma_{_{\mathcal{E}}}\quad\!\!\!\!\!q^{2} - r^{2} )}{2r^{8}},
\end{equation}
with $\mu=|q|/(2s)$. On the other hand the mapping $\rho^{2} = r^{2} - \sigma_{_{\mathcal{E}}}\quad\!\!\!\!\! q^{2}$; $\boldsymbol{d\rho} = \pm (r/\sqrt{r^{2} - \sigma_{_{\mathcal{E}}}\quad\!\!\!\!\! q^{2}})\boldsymbol{dr}$ takes the metric given in Eq.\eqref{RBH_CABH_structure_mag} to the form
%%%%%%%%%%%%%%%%%%%%%%%%
\begin{equation}\label{RBH_structure_c1_mag}
\boldsymbol{d\tilde s^{2}}
= - \left( 1 - \frac{ \mu \quad\!\!\!\!\! \rho^{2}}{\left(\rho^{2} + \sigma_{_{\mathcal{E}}} \quad\!\!\!\!\! q^{2}\right)^{3} } \right)\boldsymbol{dt^{2}} + \left( 1 - \frac{\mu \quad\!\!\!\!\! \rho^{2}}{ \left(\rho^{2} + \sigma_{_{\mathcal{E}}} \quad\!\!\!\!\! q^{2}\right)^{3} } \right)^{-1}
\boldsymbol{d\rho^{2}}
+ \rho^{2}\boldsymbol{d\Omega^{2}}
\end{equation}
 with $\rho\in[0,\infty)$. 
%%%%%%%%
Therefore, the electric dual of the pure magnetic solution (\ref{RBH_structure_c1}) is established.  

\end{itemize}

\subsection{Hybrid  Solutions, with $q_{_{\mathcal{E}}}^{2} \neq 0  \neq q_{_{\mathcal{B}}}^{2}$.}

Let us start by presenting the
general case, in which both charges are nonzero. In the $\mathcal{L}(\mathcal{F})$ representation, for a given $\mathcal{M}(r)$, the form of the Lagrangian density and its derivative as a function of $r$ can be obtained from Eqns.\eqref{eq1} and \eqref{eq2} in Appendix \ref{AppendixA}. They are given by
%%%%
\begin{eqnarray}\label{L_NLED}
&&\mathcal{L}(r) = \frac{ r^{2}\left(r^{2} - a^{2}\right)^{\!\!^{2}}\mathcal{M}'' + r\left(r^{2} - a^{2}\right)\left(7a^{2} + 2r^{2}\right)\mathcal{M}' + 3a^{2}\left(5a^{2} - r^{2}\right)\mathcal{M} }{4r^{7}} \nonumber\\
&&\hspace{1cm}\pm \hspace{0.1cm}\frac{1}{4}\hspace{0.1cm}\sqrt{ -\frac{16q_{_{\mathcal{E}}}^{2}q_{_{\mathcal{B}}}^{2} }{ \left(r^{2} - a^{2}\right)^{\!\!^{4}} } + \frac{ \left(r^{2} - a^{2}\right)^{^{2}} \Big[ r^{2}\left(a^{2} - r^{2}\right)\mathcal{M}'' + r\left(2r^{2} - 7a^{2}\right)\mathcal{M}' + 15a^{2}\mathcal{M} \Big]^{^{2}} }{r^{14}} } 
\\
\nonumber\\
&&\mathcal{L}_{_{\mathcal{F}}}(r) = \frac{(r^{2}-a^{2})^{^{2}}}{2q_{_{\mathcal{B}}}^{2}} 
\left[2\mathcal{L}(r) +  \frac{ - r^{2}(r^{2}-a^{2})^{^{2}} \hspace{0.05cm} \mathcal{M}'' - 7a^{2}\hspace{0.05cm}r\hspace{0.05cm}(r^{2}-a^{2}) \mathcal{M}'  +  3a^{2}(3r^{2}-5a^{2})\mathcal{M} }{r^{7}} \right],
\end{eqnarray}
with $ a^{2} = \sigma_{_{\mathcal{E}}} q_{_{\mathcal{E}}}^{2} + \sigma_{_{\mathcal{B}}} q_{_{\mathcal{B}}}^{2} $. The function $\mathcal{L}(\mathcal{F})$ could in principle be obtained from Eqns.\eqref{L_NLED} and \eqref{F_invariant}.

Let us present next two applications of the hybrid case.
%%%%%%%%%%%%%%%%%%%%%%%%%%%%%%%%%%%%%%%%%%%%%%%%%%%%%%%%%%%%%%%%%%%%%%%%%%%%%%%%%%%%%%%%%%%%%%%%%%%%%%%%%%%%%%%%%%%
\subsubsection*{ {\bf Application 1} }

\begin{itemize}

\item {\it Removal of the singularity of the Reissner-Nordstr\"{o}m black hole  - hybrid case in the $\mathcal{L}(\mathcal{F})$ representation:} $-g_{_{tt}}^{} = 1/g_{_{rr}}^{} = 1 - 2m/r + q^{2}_{_{\mathcal{B}}}/r^{2}$, with $m$, $q_{_{\mathcal{B}}}$ real constants. 
%%%  
In this case, the line element in Eq.(\ref{RBH_structure}) for $a^{2}=\sigma_{_{\mathcal{E}}}\quad\!\!\!\!\!q^{2}_{_{\mathcal{E}}}$, takes the form
\begin{equation}
\boldsymbol{d\tilde s^{2}} = - \left[ 1 - \left(\frac{2m}{r}-\frac{q^{2}_{_{\mathcal{B}}}}{r^{2}}\right)\!\!\left( 1 - \frac{ \sigma_{_{\mathcal{E}}} \quad\!\!\!\!\!q^{2}_{_{\mathcal{E}}}}{r^{2}} \right) \right]\boldsymbol{dt^{2}} + \frac{\boldsymbol{dr^{2}}}{\Big( 1 - \frac{ \sigma_{_{\mathcal{E}}} \quad\!\!\!\!\!\!\!q^{2}_{_{\mathcal{E}}}}{r^{2}} \Big)\!\left[ 1 - \left(\frac{2m}{r}-\frac{q^{2}_{_{\mathcal{B}}}}{r^{2}}\right)\!\!\Big( 1 - \frac{ \sigma_{_{\mathcal{E}}} \quad\!\!\!\!\!\!\!
q^{2}_{_{\mathcal{E}}} }{r^{2}}\Big) \right] }
+ \left(r^{2} -  \sigma_{_{\mathcal{E}}} \quad\!\!\!\!\! q^{2}_{_{\mathcal{E}}}\right)\boldsymbol{d\Omega^{2}}.  \end{equation}

In this case, the seed metric has a magnetic charge, and the procedure introduces an electric charge.
The NLED model for which this line element is an exact dyonic solution of the relevant field equations 
follows from Eq.\eqref{L_NLED}, and is
implicitly given by 
%%%
\begin{eqnarray}
&&\mathcal{L} =  \frac{ 3\tilde{q}^{2}_{_{\mathcal{E}}}\!\left(5\tilde{q}^{2}_{_{\mathcal{E}}} m r \!-\! 4\tilde{q}^{2}_{_{\mathcal{E}}}q^{2}_{_{\mathcal{B}}} \!-\! m r^{3} \!+\! 2 q^{2}_{_{\mathcal{B}}} r^{2} \right) }{ 4 r^{8} } + \frac{1}{4}\sqrt{  \frac{ \left( r^{2} - \tilde{q}^{2}_{_{\mathcal{E}}} \right)^{2}\!\left( 15\tilde{q}^{2}_{_{\mathcal{E}}} m r - 12 \tilde{q}^{2}_{_{\mathcal{E}}} q^{2}_{_{\mathcal{B}}}  + 2q^{2}_{_{\mathcal{B}}}r^{2} \right)^{2}     }{r^{16}} - \frac{16\tilde{q}^{2}_{_{\mathcal{E}}}q^{2}_{_{\mathcal{B}}} }{  \left( r^{2} \!-\! \tilde{q}^{2}_{_{\mathcal{E}}} \right)^{4}  } }, \label{L_Reiss}\\
\nonumber\\
&&\mathcal{L}_{_{\mathcal{F}}}\!\!=\!\frac{ \left(r^{2}\!-\!\tilde{q}^{2}_{_{\mathcal{E}}}\right)^{^{\!\!3}}\!\!\left( 15\tilde{q}^{2}_{_{\mathcal{E}}}mr\!-\!12 \tilde{q}^{2}_{_{\mathcal{E}}}q^{2}_{_{\mathcal{B}}}\!+\! 2q^{2}_{_{\mathcal{B}}}r^{2}\right)}{4 q^{2}_{_{\mathcal{B}}} r^{8}} + \frac{1}{4}\sqrt{ \!\frac{ \left(r^{2}-\tilde{q}^{2}_{_{\mathcal{E}}}\right)^{^{\!\!6}}\!\!\left(15\tilde{q}^{2}_{_{\mathcal{E}}} m r \!-\!12 \tilde{q}^{2}_{_{\mathcal{E}}}q^{2}_{_{\mathcal{B}}}\!+\!2q^{2}_{_{\mathcal{B}}}r^{2} \right)^{^{\!\!2}} }{ q^{4}_{_{\mathcal{B}}} r^{^{\!16}} } \!-\! \frac{
16\tilde{q}^{2}_{_{\mathcal{E}}}}{q^{2}_{_{\mathcal{B}}}} 
},\label{Lf_Reiss}\\
\nonumber\\
&&\mathcal{F}\!=\!-\frac{8\tilde{q}^{2}_{_{\mathcal{E}}} q^{4}_{_{\mathcal{B}}} r^{16} }{ \left(r^{2}\!-\!\tilde{q}^{2}_{_{\mathcal{E}}}\right)^{^{\!\!2}}\!\!\left[\! \left(r^{2}\!-\!\tilde{q}^{2}_{_{\mathcal{E}}}\right)^{^{\!\!3}}\!\left( 15\tilde{q}^{2}_{_{\mathcal{E}}}mr - 12 \tilde{q}^{2}_{_{\mathcal{E}}}q^{2}_{_{\mathcal{B}}} + 2q^{2}_{_{\mathcal{B}}}r^{2}\right)  \!+\!
\sqrt{ \! \left(r^{2}-\tilde{q}^{2}_{_{\mathcal{E}}}\right)^{^{\!\!6}}\!\left(15\tilde{q}^{2}_{_{\mathcal{E}}} m r - 12 \tilde{q}^{2}_{_{\mathcal{E}}}q^{2}_{_{\mathcal{B}}} + 2q^{2}_{_{\mathcal{B}}}r^{2} \right)^{^{\!\!2}} \!-\! 16\tilde{q}^{2}_{_{\mathcal{E}}}q^{2}_{_{\mathcal{B}}}r^{^{\!16}}   } \quad\!\!\!\!\right]^{^{\!\!2}}  } \nonumber\\
&& \quad\quad 
 + \quad\!\!\! \frac{q^{2}_{_{\mathcal{B}}}}{2\!\left(r^{2}\!-\!\tilde{q}^{2}_{_{\mathcal{E}}}\right)^{^{\!\!2}}}. \label{f_Reiss} 
\end{eqnarray}
%%%%
being $\tilde{q}^{2}_{_{\mathcal{E}}} = \sigma_{_{\mathcal{E}}} q^{2}_{_{\mathcal{E}}}$. The mapping $\rho^{2} =  r^{2} - \tilde{q}^{2}_{_{\mathcal{E}}}$; $\boldsymbol{d\rho} = \pm (r/\sqrt{r^{2} - \tilde{q}^{2}_{_{\mathcal{E}}} })\boldsymbol{dr}$ yields
%%%%%
\begin{equation}\label{Hibrido_LF}
\boldsymbol{d\tilde s^{2}} = - \left( 1 - \frac{2 \quad\!\!\!\!\! m \quad\!\!\!\!\! \rho^{2} }{ \left(\rho^{2} +  \quad\!\!\!\!\! \tilde{q}^{2}_{_{\mathcal{E}}}\right)^{\frac{3}{2}} } + \frac{ q^{2}_{_{\mathcal{B}}} \quad\!\!\!\!\! \rho^{2} }{ \left(\rho^{2} +   \quad\!\!\!\!\! \tilde{q}^{2}_{_{\mathcal{E}}}\right)^{2} } \right)\boldsymbol{dt^{2}} + \left( 1 - \frac{2 \quad\!\!\!\!\! m \quad\!\!\!\!\! \rho^{2}}{ \left(\rho^{2} + \quad\!\!\!\!\! \tilde{q}^{2}_{_{\mathcal{E}}}\right)^{\frac{3}{2}} } + \frac{ q^{2}_{_{\mathcal{B}}} \quad\!\!\!\!\! \rho^{2} }{ \left(\rho^{2} +  \quad\!\!\!\!\! \tilde{q}^{2}_{_{\mathcal{E}}}\right)^{2} }\right)^{-1}
\boldsymbol{d\rho^{2}}
+ \rho^{2}\boldsymbol{d\Omega^{2}} % 
\end{equation}
%%%%
This solution can be interpreted as a dyonic generalization
of the Ay\'on-Garc\'ia black hole, in which the electric field is the only responsible for the elimination of the singularity of the spacetime. In the particular case of vanishing of electric field, \emph{i.e.} if $\tilde{q}_{_{\mathcal{E}}}=0$, the line element (\ref{Hibrido_LF}) becomes the magnetically charged Reissner-Nordstr\"{o}m black hole solution. 

%%%%
The behaviour for weak fields follows from
Eq.(\ref{f_Reiss}): 
\begin{equation}
\mathcal{F}  \sim  -\frac{2 \tilde{q}^{2}_{_{\mathcal{E}}} q^{4}_{_{\mathcal{B}}} }{\left(   q^{2}_{_{\mathcal{B}}}  +    \sqrt{ q^{4}_{_{\mathcal{B}}} - 4q^{2}_{_{\mathcal{B}}}\tilde{q}^{2}_{_{\mathcal{E}}}  }  \right)^{2}  r^{4}  } + \frac{q^{2}_{_{\mathcal{B}}}}{2r^{4}}  
\sim \frac{   q^{2}_{_{\mathcal{B}}}  }{\left( q^{2}_{_{\mathcal{B}}} + \sqrt{ q^{4}_{_{\mathcal{B}}} - 4\tilde{q}^{2}_{_{\mathcal{E}}}q^{2}_{_{\mathcal{B}}}  }  \right)  } \frac{ \sqrt{ q^{4}_{_{\mathcal{B}}} - 4\tilde{q}^{2}_{_{\mathcal{E}}}q^{2}_{_{\mathcal{B}}} } }{ r^{4} } \quad\textup{with}\quad q^{4}_{_{\mathcal{B}}} - 4\tilde{q}^{2}_{_{\mathcal{E}}}q^{2}_{_{\mathcal{B}}} >0.
\end{equation}
From Eq.(\ref{L_Reiss}) we have that 
%%%%
\begin{equation}\label{L_assint}
\mathcal{L} \sim   \frac{  \sqrt{ q^{4}_{_{\mathcal{B}}}  - 4\tilde{q}^{2}_{_{\mathcal{E}}}q^{2}_{_{\mathcal{B}}}  } }{2r^{4}} \quad\quad \Rightarrow \quad\quad \mathcal{L} \sim  \left( \frac{ q^{2}_{_{\mathcal{B}}}  + \sqrt{ q^{4}_{_{\mathcal{B}}} - 4\tilde{q}^{2}_{_{\mathcal{E}}}q^{2}_{_{\mathcal{B}}}  } }{2q^{2}_{_{\mathcal{B}}}} \right) \quad\!\!\!\!\!\!\mathcal{F}. 
\end{equation}
The result from Eq.(\ref{Lf_Reiss}), %%%%%%
\begin{equation}
\mathcal{L}_{_{\mathcal{F}}} \sim \frac{ q^{2}_{_{\mathcal{B}}} + \sqrt{q^{4}_{_{\mathcal{B}}} - 4\tilde{q}^{2}_{_{\mathcal{E}}}q^{2}_{_{\mathcal{B}}}} }{2q^{2}_{_{\mathcal{B}}} }    
\end{equation}
is consistent with Eq.(\ref{L_assint}). However, the Maxwell limit, $\mathcal{L} \sim \mathcal{F}$ and $\mathcal{L}_{_{\mathcal{F}}} \sim 1$, will be possible only if $\tilde{q}_{_{\mathcal{E}}}=0$, in accordance with the theorem in \cite{Bronnikov2000}.
%%%%%%%%%%%%%%%%%%%%%%%%%%%%%%%%%%%%%%%%%%%%%%%%%%%%%%
\end{itemize}
%%%%%%%%%%%%%%%%%%%%%%%%%%%%%%%%%%%%%%%%%%%%%%%%%%%%%%%%%%%%%%%%%%%%%%%%%%%%%%%%%%%%%%%%%%%%%%%%%%%%%%%%%%%%%%%%%%%

In the case of the $\mathcal{H}(\mathcal{P})$ representation, the analogous expressions follow from Eqns.\eqref{eq3}
and \eqref{eq4}:
\begin{eqnarray}\label{H_NLED}
&&-\mathcal{H}(r) = \frac{ r^{2}\left(r^{2} - a^{2}\right)^{\!\!^{2}}\mathcal{M}'' + r\left(r^{2} - a^{2}\right)\left(7a^{2} + 2r^{2}\right)\mathcal{M}' + 3a^{2}\left(5a^{2} - r^{2}\right)\mathcal{M} }{4r^{7}} \nonumber\\
&&\hspace{1cm}\pm \hspace{0.1cm}\frac{1}{4}\hspace{0.1cm}\sqrt{ -\frac{16q_{_{\mathcal{E}}}^{2}q_{_{\mathcal{B}}}^{2} }{ \left(r^{2} - a^{2}\right)^{\!\!^{4}} } + \frac{ \left(r^{2} - a^{2}\right)^{^{2}} \Big[ r^{2}\left(a^{2} - r^{2}\right)\mathcal{M}'' + r\left(2r^{2} - 7a^{2}\right)\mathcal{M}' + 15a^{2}\mathcal{M} \Big]^{^{2}} }{r^{14}} } 
\\
\nonumber\\
&&\mathcal{H}_{_{\mathcal{P}}}(r) = -\frac{(r^{2}-a^{2})^{^{2}}}{2q_{_{\mathcal{E}}}^{2}} 
\left[  2\mathcal{H}(r) + \frac{ r^{2}(r^{2}-a^{2})^{^{2}} \hspace{0.05cm} \mathcal{M}''  + 7a^{2}\hspace{0.05cm}r\hspace{0.05cm}(r^{2}-a^{2}) \mathcal{M}' -  3a^{2}(3r^{2}-5a^{2})\mathcal{M} }{r^{7}} \right]
\end{eqnarray}
The function $\mathcal{H}(\mathcal{P})$ could be obtained from Eqns. \eqref{H_NLED} and \eqref{P_invariant}.
\subsubsection*{{\bf Application 2}}

\begin{itemize}
\item {\it Removal of the singularity of the Reissner-Nordstr\"{o}m black hole - hybrid case in the $\mathcal{H}(\mathcal{P})$ representation:} $\quad -g_{_{tt}}^{} = 1/g_{_{rr}}^{} = 1 - 2m/r + q^{2}_{_{\mathcal{E}}}/r^{2}$, with $m$, $q_{_{\mathcal{E}}}$ constants. 
%%%   
In this case, the line element in Eq.(\ref{RBH_structure}) for $a^{2}=\sigma_{_{\mathcal{B}}}\quad\!\!\!\!\!q^{2}_{_{\mathcal{B}}}$, takes the form
\begin{equation}
\label{rmh1}
\boldsymbol{d\tilde s^{2}}_{} = - \left[ 1 - \left(\frac{2m}{r}-\frac{q^{2}_{_{\mathcal{E}}}}{r^{2}}\right)\!\!\left( 1 - \frac{ \sigma_{_{\mathcal{B}}}\quad\!\!\!\!\!q^{2}_{_{\mathcal{B}}}}{r^{2}} \right) \right]\boldsymbol{dt^{2}} + \frac{\boldsymbol{dr^{2}}}{\Big( 1 - \frac{ \sigma_{_{\mathcal{B}}}\quad\!\!\!\!\!\!\!q^{2}_{_{\mathcal{B}}}}{r^{2}} \Big)\!\left[ 1 - \left(\frac{2m}{r}-\frac{q^{2}_{_{\mathcal{E}}}}{r^{2}}\right)\!\!\Big( 1 - \frac{ \sigma_{_{\mathcal{B}}}\quad\!\!\!\!\!\!\!q^{2}_{_{\mathcal{B}}}}{r^{2}}\Big) \right] }
+ \left(r^{2} -  \sigma_{_{\mathcal{B}}}\quad\!\!\!\!\! q^{2}_{_{\mathcal{B}}}\right)\boldsymbol{d\Omega^{2}}.  \end{equation}

Notice that the algorithm adds in this case magnetic charge to the metric. The NLED model to which this line element is associated to an exact (dyonic) solution of the relevant field equations 
follows from Eq.\eqref{H_NLED}, and
is
implicitly given by 
%%%
\begin{eqnarray}
&&\mathcal{H} =  - \frac{ 3\tilde{q}^{2}_{_{\mathcal{B}}}\!\left(5\tilde{q}^{2}_{_{\mathcal{B}}} m r \!-\! 4\tilde{q}^{2}_{_{\mathcal{B}}}q^{2}_{_{\mathcal{E}}} \!-\! m r^{3} \!+\! 2 q^{2}_{_{\mathcal{E}}} r^{2} \right) }{ 4 r^{8} } - \frac{1}{4}\sqrt{  \frac{ \left( r^{2} - \tilde{q}^{2}_{_{\mathcal{B}}} \right)^{2}\!\left( 15\tilde{q}^{2}_{_{\mathcal{B}}} m r - 12 q^{2}_{_{\mathcal{E}}} \tilde{q}^{2}_{_{\mathcal{B}}}  + 2q^{2}_{_{\mathcal{E}}}r^{2} \right)^{2}     }{r^{16}} - \frac{16q^{2}_{_{\mathcal{E}}}\tilde{q}^{2}_{_{\mathcal{B}}} }{  \left( r^{2} \!-\! \tilde{q}^{2}_{_{\mathcal{B}}} \right)^{4}  } }, \label{H_Reiss}\\
\nonumber\\
&&\mathcal{H}_{_{\mathcal{P}}}\!\!=\!\frac{ \left(r^{2}\!-\!\tilde{q}^{2}_{_{\mathcal{B}}}\right)^{^{\!\!3}}\!\!\left( 15\tilde{q}^{2}_{_{\mathcal{B}}}mr\!-\!12 \tilde{q}^{2}_{_{\mathcal{B}}}q^{2}_{_{\mathcal{E}}}\!+\! 2q^{2}_{_{\mathcal{E}}}r^{2}\right)}{4 q^{2}_{_{\mathcal{E}}} r^{8}} + \frac{1}{4}\sqrt{ \!\frac{ \left(r^{2}-\tilde{q}^{2}_{_{\mathcal{B}}}\right)^{^{\!\!6}}\!\!\left(15\tilde{q}^{2}_{_{\mathcal{B}}} m r \!-\!12 \tilde{q}^{2}_{_{\mathcal{B}}}q^{2}_{_{\mathcal{E}}}\!+\!2q^{2}_{_{\mathcal{E}}}r^{2} \right)^{^{\!\!2}} }{ q^{4}_{_{\mathcal{E}}} r^{^{\!16}} } \!-\! \frac{
16\tilde{q}^{2}_{_{\mathcal{B}}}}{q^{2}_{_{\mathcal{E}}}} 
},\label{Hp_Reiss}\\
\nonumber\\
&&\mathcal{P}\!=\!\frac{8\tilde{q}^{2}_{_{\mathcal{B}}} q^{4}_{_{\mathcal{E}}} r^{16} }{ \left(r^{2}\!-\!\tilde{q}^{2}_{_{\mathcal{B}}}\right)^{^{\!\!2}}\!\!\left[\! \left(r^{2}\!-\!\tilde{q}^{2}_{_{\mathcal{B}}}\right)^{^{\!\!3}}\!\left( 15\tilde{q}^{2}_{_{\mathcal{B}}}mr - 12 \tilde{q}^{2}_{_{\mathcal{B}}}q^{2}_{_{\mathcal{E}}} + 2q^{2}_{_{\mathcal{E}}}r^{2}\right)  \!+\!
\sqrt{ \! \left(r^{2}-\tilde{q}^{2}_{_{\mathcal{B}}}\right)^{^{\!\!6}}\!\left(15\tilde{q}^{2}_{_{\mathcal{B}}} m r -12 \tilde{q}^{2}_{_{\mathcal{B}}}q^{2}_{_{\mathcal{E}}} + 2q^{2}_{_{\mathcal{E}}}r^{2} \right)^{^{\!\!2}} \!-\! 16q^{2}_{_{\mathcal{E}}}\tilde{q}^{2}_{_{\mathcal{B}}}r^{^{\!16}}   } \quad\!\!\!\!\right]^{^{\!\!2}}  } \nonumber\\
&& \quad\quad 
 - \quad\!\!\! \frac{q^{2}_{_{\mathcal{E}}}}{2\!\left(r^{2}\!-\!\tilde{q}^{2}_{_{\mathcal{B}}}\right)^{^{\!\!2}}}. \label{p_Reiss} 
\end{eqnarray}
%%%%
%%%%
being $\tilde{q}^{2}_{_{\mathcal{B}}} = \sigma_{_{\mathcal{B}}} q^{2}_{_{\mathcal{B}}}$. Applying the mapping $\rho^{2} =  r^{2} - \tilde{q}^{2}_{_{\mathcal{B}}}$; $\boldsymbol{d\rho} = \pm (r/\sqrt{r^{2} - \tilde{q}^{2}_{_{\mathcal{B}}} })\boldsymbol{dr}$
to Eq.\eqref{rmh1},
we obtain
%%%%%
\begin{equation}\label{Hibrido_HP}
\boldsymbol{ds^{2}} = - \left( 1 - \frac{2 \quad\!\!\!\!\! m \quad\!\!\!\!\! \rho^{2} }{ \left(\rho^{2} +  \quad\!\!\!\!\! \tilde{q}^{2}_{_{\mathcal{B}}}\right)^{\frac{3}{2}} } + \frac{ q^{2}_{_{\mathcal{E}}} \quad\!\!\!\!\! \rho^{2} }{ \left(\rho^{2} +   \quad\!\!\!\!\! \tilde{q}^{2}_{_{\mathcal{B}}}\right)^{2} } \right)\boldsymbol{dt^{2}} + \left( 1 - \frac{2 \quad\!\!\!\!\! m \quad\!\!\!\!\! \rho^{2}}{ \left(\rho^{2} + \quad\!\!\!\!\! \tilde{q}^{2}_{_{\mathcal{B}}}\right)^{\frac{3}{2}} } + \frac{ q^{2}_{_{\mathcal{E}}} \quad\!\!\!\!\! \rho^{2} }{ \left(\rho^{2} +  \quad\!\!\!\!\! \tilde{q}^{2}_{_{\mathcal{B}}}\right)^{2} }\right)^{-1}
\boldsymbol{d\rho^{2}}
+ \rho^{2}\boldsymbol{d\Omega^{2}} % 
\end{equation}
%%%%
This solution can be interpreted as a dyonic Ay\'on-Garc\'ia black hole generalization, in which the magnetic field is the sole responsible for the elimination of the singuarity. Thus, in the particular case of vanishing of magnetic field, \emph{i.e.} if $\tilde{q}_{_{\mathcal{B}}}=0$, the line element (\ref{Hibrido_HP}) becomes the electrically charged Reissner-Nordstr\"{o}m black hole.  

%%%%
On the other hand, the behaviour for weak fields can be obtained 
from Eq.(\ref{p_Reiss}),  yielding
\begin{equation}
\mathcal{P}  \sim  \frac{2 \tilde{q}^{2}_{_{\mathcal{B}}} q^{4}_{_{\mathcal{E}}} }{\left(   q^{2}_{_{\mathcal{E}}}  +    \sqrt{ q^{4}_{_{\mathcal{E}}} - 4q^{2}_{_{\mathcal{E}}}\tilde{q}^{2}_{_{\mathcal{B}}}  }  \right)^{2}  r^{4}  } -\frac{q^{2}_{_{\mathcal{E}}}}{2r^{4}}  
\sim -\frac{   q^{2}_{_{\mathcal{E}}}  }{\left( q^{2}_{_{\mathcal{E}}} + \sqrt{ q^{4}_{_{\mathcal{E}}} - 4q^{2}_{_{\mathcal{E}}}\tilde{q}^{2}_{_{\mathcal{B}}}  }  \right)  } \frac{ \sqrt{ q^{4}_{_{\mathcal{E}}} - 4q^{2}_{_{\mathcal{E}}}\tilde{q}^{2}_{_{\mathcal{B}}} } }{ r^{4} } \quad\textup{with}\quad q^{4}_{_{\mathcal{E}}} - 4q^{2}_{_{\mathcal{E}}}\tilde{q}^{2}_{_{\mathcal{B}}} >0.
\end{equation}
From Eq.(\ref{H_Reiss}) it follows that
%%%%
\begin{equation}\label{H_assint}
\mathcal{H} \sim  - \frac{  \sqrt{ q^{4}_{_{\mathcal{E}}}  - 4q^{2}_{_{\mathcal{E}}}\tilde{q}^{2}_{_{\mathcal{B}}}  } }{2r^{4}} \quad\quad \Rightarrow \quad\quad \mathcal{H} \sim  \left( \frac{ q^{2}_{_{\mathcal{E}}}  + \sqrt{ q^{4}_{_{\mathcal{E}}} - 4q^{2}_{_{\mathcal{E}}}\tilde{q}^{2}_{_{\mathcal{B}}}  } }{2q^{2}_{_{\mathcal{E}}}} \right) \quad\!\!\!\!\!\!\mathcal{P}. 
\end{equation}
The result from Eq.(\ref{Hp_Reiss}), namely,
%%%%%%
\begin{equation}
\mathcal{H}_{_{\mathcal{P}}} \sim \frac{ q^{2}_{_{\mathcal{E}}} + \sqrt{q^{4}_{_{\mathcal{E}}} - 4q^{2}_{_{\mathcal{E}}}\tilde{q}^{2}_{_{\mathcal{B}}}} }{2q^{2}_{_{\mathcal{E}}} }    
\end{equation}
is consistent with Eq.(\ref{H_assint}). Notice that the Maxwell limit, $\mathcal{H} \sim \mathcal{P}$ and $\mathcal{H}_{_{\mathcal{P}}} \sim 1$, will be possible only if $\tilde{q}_{_{\mathcal{B}}}=0$.
This result points
to the existence of a theorem dual to the one presented in \cite{Bronnikov2000}.
\end{itemize}
%%%%
%%
%%%%%%%%%%%%%%%%%%%%%%%%%%%%%%%%%%%%%%%%%%%%%%%%%%%%%%%%%%%%%%%%%%%%%%%%%%%%%%%%%%%%%%%%%%%%%%%%%%

\section{Conclusion}
\label{concl}
We have presented a method to obtain regular SSS-AF BH solutions of the system GR-NLED from singular SSS-AF BH solutions, which are not necessarily  vacuum or electro-vacuum solutions of GR. In particular, we have shown that the known regular BH solutions of GR-NLED
actually follow 
from the application of our method. 
A regular BH solution of GR-NLED
was obtained from the canonical acoustic black hole, which is not a solution of Einstein's equations. 
All the solutions obtained here are in accordance with the theorem presented in \cite{Bronnikov2000}. In fact, our results suggest that it should be possible to formulate a theorem dual to that one, in the ${\cal H}$ frame.

It is important to stress that the method presented here can in principle be used to remove the singularity present in BH solutions in the context of other nonlinear theories, for instance those including scalar fields. 
Finally, it is worth mentioning that the region $r<|a|$ (in which the metric has Euclidean signature), which seems to have been excluded from spacetime by the  procedure, could be studied along the lines presented in 
\cite{Hellaby1997}. We hope to return to these points in a future publication.

%%%%%%%%%%%%%%%%%%%%%%%%%%%%%%%%%%%%%%%%%%%%%%%%%%%%%%%%%%%%%%%%%%%%%%%%%%%%%%%%%%%%%%%%%%%%%%%%%%%%%%%%%%%
\appendix \section{GR-NLED field equations}\label{AppendixA}
Some useful expressions are displayed next.
%%%%%%%%%%%%%%%%%
For the  regular black hole geometries with metric given in Eq.(\ref{RBH_structure}), the non-null components of the Einstein tensor are given by
\begin{eqnarray}
&&
G_{t}{}^{t} = G_{r}{}^{r} = \frac{2(a^{2} - r^{2})\mathcal{M}'}{ r^{4}} - \frac{6a^{2}\mathcal{M}}{ r^{5}}, \label{Gtt_Grr}\\ 
&&
G_{\theta}{}^{\theta}=G_{\varphi}{}^{\varphi} = -\frac{\left( r^{2} - a^{2}\right)^{2}\mathcal{M}''}{ r^{5}} + \frac{7a^{2}\left( a^{2} - r^{2}\right)\mathcal{M}'}{ r^{6}} + \frac{3a^{2}\left( 3r^{2} - 5a^{2}\right)\mathcal{M}}{ r^{7}}. \label{Gtete_Gphiphi} 
\end{eqnarray}
The 
GR-NLED field equations in the $\mathcal{L}(\mathcal{F})$ representation are
\begin{eqnarray}
&&{C}_{t}{}^{t} = {C}_{r}{}^{r} = 0 \quad\Rightarrow\quad
\frac{2(a^{2} - r^{2})\mathcal{M}'}{ r^{4}} - \frac{6a^{2}\mathcal{M}}{ r^{5}}
=  -\frac{2q_{_{\mathcal{E}}}^{2} }{(r^{2}-a^{2})^{^{2}}\mathcal{L}_{\mathcal{F}}} - 2\mathcal{L},
\label{eq1}\\ 
%\nonumber\\
%\nonumber\\
&&
{C}_{\theta}{}^{\theta} = {C}_{\varphi}{}^{\varphi} = 0 \quad\Rightarrow\quad
-\frac{\left( r^{2} - a^{2}\right)^{2}\mathcal{M}''}{ r^{5}} + \frac{7a^{2}\left( a^{2} - r^{2}\right)\mathcal{M}'}{ r^{6}} + \frac{3a^{2}\left( 3r^{2} - 5a^{2}\right)\mathcal{M}}{ r^{7}}
= \frac{2q_{_{\mathcal{B}}}^{2}\mathcal{L}_{\mathcal{F}}}{(r^{2}-a^{2})^{^{2}}} - 2\mathcal{L}.
\label{eq2}
\end{eqnarray}
\\
%%%%%%%%%%%%%%%%%%%%%%%%%%%%%%%%%%%%%
The GR-NLED field equations in the $\mathcal{H}(\mathcal{P})$ representation are
\begin{eqnarray}
&&{C}_{t}{}^{t} = {C}_{r}{}^{r} = 0 \quad\Rightarrow\quad
\frac{2(a^{2} - r^{2})\mathcal{M}'}{ r^{4}} - \frac{6a^{2}\mathcal{M}}{ r^{5}} = 
-\frac{2q^{2}_{_{\mathcal{B}}}}{(r^{2}-a^{2})^{2}\mathcal{H}_{\mathcal{P}}}+2\mathcal{H},\label{eq3}\\ 
%\nonumber\\
&&
{C}_{\theta}{}^{\theta} = {C}_{\varphi}{}^{\varphi} = 0 \quad\Rightarrow\quad
-\frac{\left( r^{2} - a^{2}\right)^{2}\mathcal{M}''}{ r^{5}} + \frac{7a^{2}\left( a^{2} - r^{2}\right)\mathcal{M}'}{ r^{6}} + \frac{3a^{2}\left( 3r^{2} - 5a^{2}\right)\mathcal{M}}{ r^{7}} =
\frac{2q^{2}_{_{\mathcal{E}}}\mathcal{H}_{\mathcal{P}}}{(r^{2}-a^{2})^{2}}+2\mathcal{H} .
\label{eq4}
\end{eqnarray}
%%%%%%%%%%%%%%%%%%%
\section{Curvature invariants}\label{AppendixB}
Some curvature invariants for the metric given in Eq.(\ref{RBH_structure}):

\begin{eqnarray}
&&R = \frac{2\left(r^{2}-a^{2}\right)^{2}}{r^{5}}\mathcal{M}'' + \frac{2\left(r^{2}-a^{2}\right)\left(7a^{2}+2r^{2}\right)}{r^{6}} \mathcal{M}'  + \frac{6a^{2}\left(5a^{2}-r^{2}\right) }{r^{7}}\mathcal{M}, \label{R_BH_gen1}\\
\nonumber\\
%%%%%
&&R_{\alpha\beta}R^{\alpha\beta} =\frac{2}{r^{14}} \Big\{ r^{2}\left(a^{2}-r^{2}\right)^{2} \left[ r^{2}\left(a^{2}-r^{2}\right)^{2}\mathcal{M}'' - 14a^{2}r\left(a^{2}-r^{2}\right)\mathcal{M}' + 6 a^{2}\left(5a^{2}-3r^{2}\right)\mathcal{M}  \right] \mathcal{M}''    \nonumber\\
&& \quad\quad\quad\quad\quad\quad\quad + \!\!\quad\!\! r\left(a^{2}-r^{2}\right)\left[ r\left(49a^{4}+4r^{4}\right)\left(a^{2}-r^{2}\right)\mathcal{M}' - 6a^{2}\left(35a^{4}-21a^{2}r^{2}+4r^{4}\right) \mathcal{M} \right]\mathcal{M}'  \nonumber\\
&& \quad\quad\quad\quad\quad\quad\quad + \!\!\quad\!\! 9a^{4}\left(25a^{4}-30a^{2}r^{2}+13r^{4}\right)\mathcal{M}^{2}   \Big\}, \label{R_BH_gen2}
\\
\nonumber\\
%%%%%
&&R_{\alpha\beta\mu\nu}R^{\alpha\beta\mu\nu} \!=\!\frac{4}{r^{14}} \Big\{ r^{2}(r^{2}-a^{2})^{^{2}}\left[ r^{2}(r^{2}-a^{2})^{^{2}} \mathcal{M}'' - 2r(2r^{2}-7a^{2})(r^{2}-a^{2})\mathcal{M}' + 2 (15a^{4}-15a^{2}r^{2} + 2r^{4})\mathcal{M}\right]\mathcal{M}'' \nonumber\\
&& \quad\quad\quad\quad\quad\quad\!\! + \!\!\quad\!\! r\left(a^{2}-r^{2}\right)\left[ r(a^{2} - r^{2})(49a^{4}-28a^{2}r^{2} + 8r^{4})\mathcal{M}' - 2\left(105a^{6}-135a^{4}r^{2} + 56 a^{2}r^{4} -8r^{6}\right)\mathcal{M}  
\right]\mathcal{M}' \nonumber \\
&& \quad\quad\quad\quad\quad\quad\!\! + \!\!\quad\!\! 3 (75a^{8} - 150a^{6}r^{2} + 107 a^{4}r^{4} - 28a^{2}r^{6} + 4r^{8})\mathcal{M}^{2}
\Big\}.
\label{R_BH_gen3}
\end{eqnarray}

%%%%%%%%%%%%%%%%%%%%%%%
\section{Setting the $\mathcal{M}(r)$ function by the energy conditions}\label{AppendixC}
%

%%%%%%
In GR without cosmological constant, $G_{\alpha}{}^{\beta} = 8\pi T_{\alpha}{}^{\beta}$, the Schwarzschild metric is the only spherically symmetric solution of the vacuum ($T_{\alpha}{}^{\beta}=0$) field equations. This follows from Birkhoff's theorem which states that the static, asymptotically flat solution is described only in terms of one single conserved quantity measured at infinity, namely the ADM mass \cite{ADMmass}. Hence, in order for the line element given in Eq. (\ref{RBH_structure}) to be a nontrivial solution of the Einstein field equations, the contribution of a matter field coupled to such equations is needed. 
Next we show the conditions under which
the energy-momentum tensor associated with
such matter field satisfies the weak energy condition (WEC) everywhere, for the line element (\ref{RBH_structure}).
%%%%%%%%%%%%%%%%%%%%%%%%%%%%%%%%%%%%%%%%%%%%%%%%%%%%%%%%%%%%%%%%%%%%%%%%%%%%%%%%%%%%%%%%%%%%%%%%%%%%%%%%%%%%%%
The WEC states that for any timelike vector $\boldsymbol{k} = k^{\mu}\partial_{\mu}$, with  $k_{\mu}k^{\mu}<0$, the energy-momentum tensor $T_{\mu\nu}$ obeys the inequality
$T_{\mu\nu}k^{\mu}k^{\nu} \geq 0$, which means that the local energy density $\rho_{\!_{_{loc}}}= T_{\mu\nu}k^{\mu}k^{\nu}$ as measured by any observer with timelike vector $\boldsymbol{k}$ is non-negative. Following \cite{WEC}, a diagonal energy-momentum tensor $(T_{\alpha\beta})={\rm diag} \left( T_{tt},T_{rr},T_{\theta\theta},T_{\varphi\varphi} \right)$, can be
conveniently written as follows:
%%%%%%%%%%%%%%%%%%%%%%%%%%%%%%%%%%%%%
\begin{equation}\label{diagonalEab}
T_{\alpha}{}^{\beta} = - \rho_{_{0}} \hskip.05cm \delta_{\alpha}{}^{t}\delta_{t}{}^{\beta} + P_{r} \hskip.05cm \delta_{\alpha}{}^{r}\delta_{r}{}^{\beta} + P_{\theta} \hskip.05cm \delta_{\alpha}{}^{\theta}\delta_{\theta}{}^{\beta} + P_{\varphi} \hskip.05cm \delta_{\alpha}{}^{\varphi}\delta_{\varphi}{}^{\beta} ,
\end{equation}
where $\rho_{_{0}}$ is the rest energy density of the matter, while 
$P_{r}$, $P_{\theta}$ and $P_{\varphi}$ are respectively the principal pressures along the $r$, $\theta$ and $\varphi$ directions.

The WEC leads to
\begin{equation}\label{WEC}
\rho_{_{0}} = - T_{t}{}^{t} \geq0, \quad \rho_{0} + P_{a} \geq0, \quad  a = \{ r, \theta, \varphi \}.
\end{equation}
%%%%%%%%%%%%%%%%%%%
%%%%%%%%%%%%%%%%%%%%%%%%%%%%%%%%%%%%%%%%%%%%%%%%%%%%%%%%%%%%%%%%%%%%%%%%%%%%%%%%%%%%%%%%%%%%%%%%%%%%%%%%%%%%%%
According to GR, the matter distribution, in the spacetime equipped with the metric (\ref{RBH_structure}),
is related to the metric by the Einstein field equations $T_{\alpha}{}^{\beta}=G_{\alpha}{}^{\beta}/(8\pi)$. Hence, for the line element (\ref{RBH_structure}) in the GR context, the non-zero $T_{\alpha}{}^{\beta}$ components are (see equations (\ref{Gtt_Grr})-(\ref{Gtete_Gphiphi}) of Appendix \ref{AppendixA}):
\begin{eqnarray}
&&8\pi T_{t}{}^{t} = 8\pi T_{r}{}^{r} = \frac{2(a^{2} - r^{2})\mathcal{M}'}{ r^{4}} - \frac{6a^{2}\mathcal{M}}{ r^{5}}, \\
&& 8\pi T_{\theta}^{\theta} = 8\pi T_{\varphi}^{\varphi} = -\frac{\left( r^{2} - a^{2}\right)^{2}\mathcal{M}''}{ r^{5}} + \frac{7a^{2}\left( a^{2} - r^{2}\right)\mathcal{M}'}{ r^{6}} + \frac{3a^{2}\left( 3r^{2} - 5a^{2}\right)\mathcal{M}}{ r^{7}}.
\end{eqnarray}
Therefore, by identification with Eq.(\ref{diagonalEab}), it follows that $\rho_{_{0}} + P_{r} =0$ together with,
%%%%%%%%%%%%%%%%%%%%%%%%%%%%
\begin{eqnarray}
&&\rho_{_{0}} = - T_{t}{}^{t} = \frac{1}{8\pi}\left[\frac{2(r^{2} - a^{2})\mathcal{M}'}{ r^{4}} + \frac{6a^{2}\mathcal{M}}{ r^{5}}\right],\label{V_WECp} \\
&&\rho_{_{0}} + P_{\theta} = \rho_{0} + P_{\varphi} = \frac{1}{8\pi}\left[ -\frac{(r^{2}-a^{2})^{^{2}}\mathcal{M}''}{r^{5}} + \frac{2(r^{2}-a^{2})^{^{2}}\mathcal{M}'}{r^{6}} + \frac{5a^{2}(r^{2}-a^{2})}{r^{6}}\left(\frac{3\mathcal{M}}{r} - \mathcal{M}' \right)\right]. \label{V_WECpP}
\end{eqnarray}
%%%%%%%%%%%%%%%%%%%%%%%%%%%%%%%
The expressions (\ref{V_WECp})-(\ref{V_WECpP}) are positive for any value of the radial coordinate $r\in[|a|,\infty)$ %%%
if the $\mathcal{M}(r)$ is such that, 
\begin{equation}\label{WEC_1}
\mathcal{M}\geq0, \quad\quad\quad \mathcal{M}'\geq0, \quad\quad\quad \mathcal{M}''\leq0, \quad\quad\quad \textup{and} \quad\quad\quad \frac{3\mathcal{M}}{r} - \mathcal{M}'\geq0 \quad\quad\quad \forall r\geq|a|,   
\end{equation}
which implies that the weak energy condition (\ref{WEC}) and the null energy condition\footnote{The NEC states that for any null vector, $n^{\alpha}$, $T_{\mu\nu}n^{\mu}n^{\nu}\geq0$. In terms of (\ref{diagonalEab}) the NEC implies: $\rho_{_{0}} + P_{a} \geq0$,  $a = \{ r, \theta, \varphi \}$} (NEC), are fulfilled  
everywhere.

%%%%%%%%%%%%%%%%%%%%%%%%%%%%%%%%%%%%%%%%%%%%%%%%%%%%%%%%%%%%%%%%%%%%%%%%%%%%%%%%%%%%%%%%%%%%%%%%%%%%%%%%%%%

\textbf{Acknowledgments}:
P.M.C.C. was partially supported by Coordena\c c\~ao de Aperfei\c coamento de Pessoal de N\'ivel Superior-Brasil (CAPES) - C\'odigo de Financiamento 001.

\end{document}